\documentclass[a4paper]{article}
\usepackage[margin=1in]{geometry}
\usepackage{graphicx} 
\usepackage{amsmath}
\usepackage{amssymb}
\usepackage{float}
\usepackage{multirow}
\usepackage[table,xcdraw]{xcolor}
\usepackage{longtable}
\usepackage{authblk}
\usepackage{url}

\title{Investigating Quantum Feature Maps in Quantum Support Vector Machines for Lung Cancer Classification}
\author[1]{A. Toufah}
\author[1]{M. A. Kadim}
\author[1]{My Y. El Hafidi\thanks{youssef.elhafidi@univh2c.ma}}
\affil[1]{Quantum Physics and Spintronics Team, Condensed Matter Physics Laboratory (LPMC), Faculty of Sciences Ben M'Sik, Hassan II University of Casablanca, Morocco}
\date{April 2025}

\begin{document}

\maketitle

\begin{abstract}
In recent years, quantum machine learning has emerged as a promising intersection between quantum physics and artificial intelligence, particularly in domains requiring advanced pattern recognition such as healthcare. This study investigates the effectiveness of Quantum Support Vector Machines (QSVM), which leverage quantum mechanical phenomena like superposition and entanglement to construct high-dimensional Hilbert spaces for data classification. Focusing on lung cancer diagnosis—a concrete and critical healthcare application—we analyze how different quantum feature maps influence classification performance. Using a real-world dataset of 309 patient records with significant class imbalance (39 non-cancer vs. 270 cancer cases), we constructed six balanced subsets for robust evaluation. QSVM models were implemented using Qiskit and executed on the \texttt{qasm\_simulator}, employing three distinct quantum feature maps: ZFeatureMap, ZZFeatureMap, and PauliFeatureMap. Performance was assessed using accuracy, precision, recall, specificity, and F1-score. Results show that the PauliFeatureMap consistently outperformed the others, achieving perfect classification in three subsets and strong performance overall. These findings demonstrate how quantum computational principles can be harnessed to enhance diagnostic capabilities, reinforcing the importance of physics-based modeling in emerging AI applications within healthcare.\\

\end{abstract}
\section{Introduction}

Lung cancer remains one of the leading causes of cancer-related mortality worldwide, with early detection being critical to improving patient outcomes. Accurate and efficient diagnostic tools are therefore essential to assist healthcare professionals in making timely decisions. Machine learning (ML) techniques, particularly Support Vector Machines (SVMs), have demonstrated strong performance in various medical classification tasks, including cancer diagnosis \cite{cristianini2000introduction}.

Classical SVMs operate by finding an optimal hyperplane that separates data points of different classes, relying heavily on the choice of kernel functions to project data into higher-dimensional spaces where it becomes more easily separable. Despite their success, classical models may struggle to capture highly complex data structures inherent in biomedical data, motivating exploration into emerging computational paradigms.

Quantum computing offers a new framework for data processing by leveraging quantum mechanical principles such as superposition and entanglement \cite{schuld2015introduction}. Quantum Support Vector Machines (QSVMs) have been proposed as quantum-enhanced analogs of classical SVMs, aiming to exploit quantum feature spaces to achieve potentially superior classification capabilities \cite{havlivcek2019supervised}. A critical component in QSVMs is the feature map, which encodes classical data into quantum states. The design of these quantum feature maps determines how information is embedded into the quantum Hilbert space and can significantly influence the model's ability to distinguish between classes \cite{biamonte2017quantum}.

In this study, we investigate the role of different quantum feature maps in QSVM performance on a lung cancer classification task. By comparing classical SVMs with various kernel functions and QSVMs utilizing different quantum feature maps, we aim to assess how quantum data encoding impacts classification outcomes in a healthcare context.

\section{Related Work}

Support Vector Machines (SVMs) have been widely adopted in healthcare applications, particularly for disease diagnosis and classification tasks. In the context of cancer detection, SVM models have demonstrated robust performance due to their capacity to handle high-dimensional and complex data \cite{huang2002application, polat2007lung}. Several studies have reported high accuracy rates when applying SVMs to lung cancer datasets, establishing them as a reliable baseline method for clinical prediction tasks \cite{alam2018multi, parveen2014classification}.

Quantum Machine Learning (QML) has recently begun to penetrate the healthcare domain. For example, Ramos-Calderer et al. applied quantum machine learning techniques to breast cancer classification tasks, showing the potential for quantum models to perform competitively with classical approaches \cite{ramos2021quantum}. Similarly, Schuld et al. explored circuit-centric quantum classifiers, highlighting the role of data encoding strategies in determining model performance \cite{schuld2020circuit}.

A critical factor influencing QSVM performance is the selection of \textit{feature maps}, which define how classical data is embedded into quantum Hilbert spaces. Studies such as Mitarai et al. \cite{mitarai2018quantum} emphasize that appropriate feature mapping can significantly impact the classifier's ability to separate classes in complex datasets. However, there remains a lack of systematic comparison of different quantum feature maps, especially in real-world medical datasets like lung cancer diagnosis.

This gap motivates the present study, where we conduct a comparative analysis between classical SVM models and QSVM models utilizing various feature maps, applied to a lung cancer classification task.
\section{Methodology}

This section outlines the methodology used to classify lung cancer data with classical and quantum support vector machines (SVMs). We first describe the data preprocessing steps, followed by an overview of SVM and quantum SVM models. Finally, we present the evaluation metrics used to assess the models' performance.

\subsection{Dataset Preprocessing and Balancing}

The dataset used in this study was retrieved from a public Kaggle repository \cite{sendil2025lung}. It comprises 309 patient records, each annotated with a binary label indicating lung cancer diagnosis. The target variable, denoted by \( y \in \{0, 1\} \), corresponds to whether a patient has lung cancer (1) or not (0). Among the 309 samples, 39 were labeled as negative cases (\( y = 0 \)), while the remaining 270 samples were labeled as positive cases (\( y = 1 \)).

The features in the dataset are primarily binary, encoded as \texttt{Yes} or \texttt{No}, with the exception of the gender variable (\texttt{M} or \texttt{F}) and age, which is continuous. To prepare the data for model training, all categorical binary features were numerically encoded using a simple mapping: \texttt{Yes} was mapped to 1, and \texttt{No} to 0. Similarly, gender was binarized by mapping \texttt{M} to 1 and \texttt{F} to 0. The continuous feature, age, was normalized using standardization:

\begin{equation}
\hat{x} = \frac{x - \mu}{\sigma}
\end{equation}

where \( x \) is the age value, \( \mu \) is the sample mean, and \( \sigma \) is the sample standard deviation. This transformation ensures that age values have zero mean and unit variance, which improves numerical stability during training and is commonly used in machine learning \cite{jain2000data}.

No features were removed from the dataset. All variables were retained to preserve the original structure and maintain consistency across both classical and quantum models. Since the dataset was already clean and contained no missing values, no imputation or data cleaning procedures were required.

To address the issue of class imbalance, we constructed a set of six balanced subsets. Each subset includes all 39 negative samples and a unique, randomly selected set of 39 positive samples without replacement. Formally, let \( D = D_0 \cup D_1 \), where \( D_0 \) is the set of all samples such that \( y = 0 \) and \( D_1 \) is the set where \( y = 1 \). We define each balanced subset \( S_i \) as:

\begin{equation}
S_i = D_0 \cup D_1^{(i)} \quad \text{for } i = 1, 2, ..., 6
\end{equation}

where \( D_1^{(i)} \subset D_1 \) and \( |D_1^{(i)}| = 39 \), with \( D_1^{(i)} \cap D_1^{(j)} = \emptyset \) for \( i \neq j \). This subset design enables repeated evaluation of the models over different balanced configurations while avoiding data leakage and overlap. The choice of balancing method was guided by the desire to preserve data integrity while addressing class imbalance, which is crucial for accurate model performance in healthcare applications \cite{chawla2002smote}.

\subsection{Support Vector Machine (SVM)}

Support Vector Machines (SVM) are powerful supervised learning algorithms used for classification tasks, where the objective is to find a decision boundary that maximally separates data points belonging to different classes \cite{Boser1992}. In the case of binary classification, given a training set \( \{ (\mathbf{x}_i, y_i) \}_{i=1}^n \), where \( \mathbf{x}_i \in \mathbb{R}^d \) are the feature vectors and \( y_i \in \{-1, 1\} \) represent the class labels, the SVM learns a decision function of the form:

\begin{equation}
f(\mathbf{x}) = \text{sign}(\mathbf{w}^\top \mathbf{x} + b)
\end{equation}

Here, \( \mathbf{w} \) is the weight vector orthogonal to the decision hyperplane, and \( b \) is the bias term. The primary goal in SVM is to maximize the margin between the two classes, which is inversely proportional to the norm of the weight vector, i.e., \( \frac{2}{\|\mathbf{w}\|} \). This process ensures that the decision boundary lies as far as possible from the closest data points, known as support vectors, minimizing the classification error.

However, in many real-world problems, data points from the two classes are not linearly separable in the original feature space. To handle such cases, the concept of soft-margin SVM is introduced. This formulation allows for misclassifications by introducing slack variables \( \xi_i \geq 0 \), and the optimization problem becomes:

\begin{equation}
\min_{\mathbf{w}, b, \boldsymbol{\xi}} \quad \frac{1}{2} \|\mathbf{w}\|^2 + C \sum_{i=1}^{n} \xi_i
\end{equation}

subject to the constraints:

\begin{equation}
y_i(\mathbf{w}^\top \mathbf{x}_i + b) \geq 1 - \xi_i, \quad \xi_i \geq 0, \quad \forall i
\end{equation}

where \( C \) is the regularization parameter that controls the trade-off between maximizing the margin and minimizing classification error. A larger \( C \) emphasizes minimizing misclassifications, while a smaller \( C \) favors larger margins with potential misclassifications.

To further extend SVM’s ability to handle non-linearly separable data, the kernel trick is introduced \cite{Elias2023}. Instead of directly applying a non-linear transformation to the feature space, SVM computes the inner products of the transformed data points in a higher-dimensional space using a kernel function. This allows the algorithm to implicitly map the data into a higher-dimensional space, where it might become linearly separable, without having to compute the transformation explicitly. The kernel function \( K(\mathbf{x}_i, \mathbf{x}_j) \) computes the inner product between two points \( \mathbf{x}_i \) and \( \mathbf{x}_j \) in the transformed feature space, thus enabling SVM to operate in this higher-dimensional space.

The most widely used kernel functions include the linear kernel, the polynomial kernel, and the radial basis function (RBF) kernel. The linear kernel, which is the simplest form of kernel, is defined as:

\begin{equation}
K(\mathbf{x}_i, \mathbf{x}_j) = \mathbf{x}_i^\top \mathbf{x}_j
\end{equation}

This kernel is particularly effective when the data is already linearly separable in the original feature space, as it does not perform any mapping but instead directly computes the inner product between the feature vectors. 

In cases where the data exhibits non-linear relationships, the polynomial kernel is often used. This kernel maps the data into a higher-dimensional space where a polynomial decision boundary can separate the classes \cite{Pontil1998}. The polynomial kernel is given by:

\begin{equation}
K(\mathbf{x}_i, \mathbf{x}_j) = (\mathbf{x}_i^\top \mathbf{x}_j + c)^d
\end{equation}

where \( c \) is a constant, and \( d \) is the degree of the polynomial. By adjusting the degree \( d \), the polynomial kernel can represent different types of decision boundaries, from simple linear boundaries (when \( d = 1 \)) to more complex, curved boundaries (as \( d \) increases).

The radial basis function (RBF) kernel, also known as the Gaussian kernel, is one of the most commonly used kernels due to its flexibility in capturing complex, non-linear relationships between data points \cite{Scholkopf2002}. The RBF kernel is defined as:

\begin{equation}
K(\mathbf{x}_i, \mathbf{x}_j) = \exp\left(-\gamma \|\mathbf{x}_i - \mathbf{x}_j\|^2\right)
\end{equation}

where \( \gamma \) is a parameter that controls the spread of the Gaussian function. The RBF kernel’s ability to assign a higher weight to nearby data points and a lower weight to distant points makes it especially suitable for problems with complex decision boundaries. It is particularly effective when the data lies in a high-dimensional, non-linear feature space and can capture intricate relationships between the input features.

By using these kernel functions, SVM can classify data even in cases where the decision boundary is non-linear. The flexibility of the kernel trick is one of the reasons for SVM’s widespread success in various machine learning applications, including text classification, image recognition, and bioinformatics \cite{Cortes1995}. Moreover, the ability to tune parameters like \( C \) and \( \gamma \) allows the model to adapt to different types of data and classification tasks \cite{Scholkopf2002}.

In this study, the RBF kernel was chosen due to its ability to handle non-linear data and the fact that no prior knowledge about the data distribution was available. The performance of the SVM was evaluated using various configurations of \( C \) and \( \gamma \), optimizing these hyperparameters to achieve the best possible model performance.

\begin{figure}[h]
    \centering
    \includegraphics[width=0.6\textwidth]{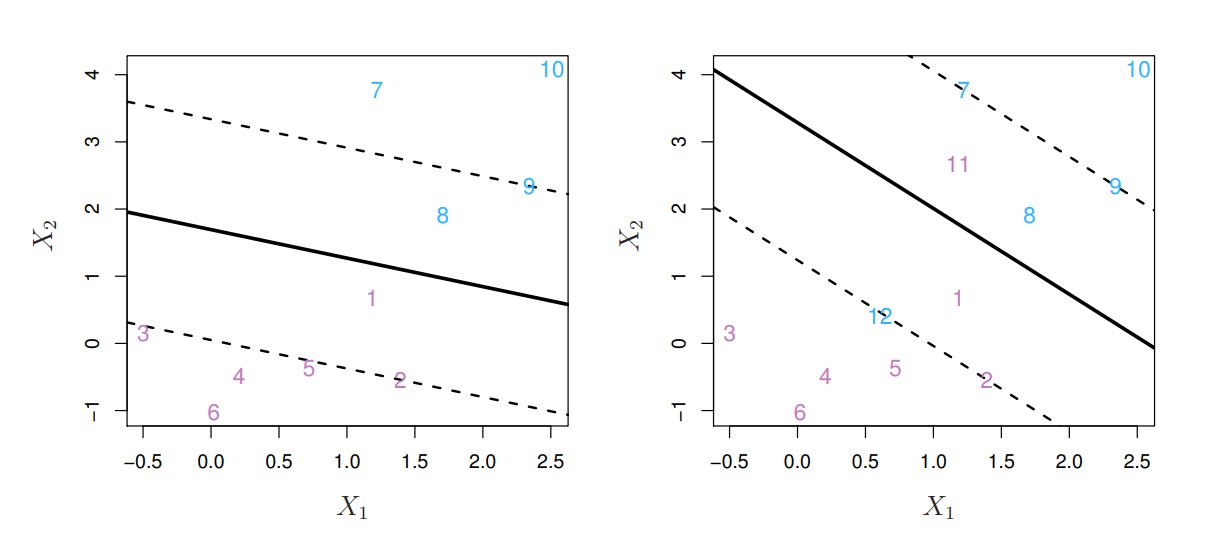}
    \caption{Illustration of a soft-margin SVM with support vectors, margin boundaries, and decision hyperplane.}
    \label{fig:svm_margin}
\end{figure}

An illustration of the SVM decision boundary with support vectors is shown in Figure~\ref{fig:svm_margin}, which visually depicts the decision hyperplane, margin boundaries, and support vectors in a two-dimensional space.

\subsection{Quantum Support Vector Machines}

Quantum Support Vector Machines (QSVM) extend the classical SVM framework by leveraging the mathematical principles of quantum mechanics—particularly, quantum Hilbert spaces and state fidelity—to perform classification tasks. The key idea behind QSVM is to encode classical input data into quantum states and use quantum circuits to compute a kernel that captures similarities between these states.

In classical SVMs, the decision function depends on the computation of an inner product (or kernel) between input vectors. In QSVM, this inner product is replaced by the quantum fidelity between two quantum states:
\begin{equation}
    K_q(x_i, x_j) = |\langle \psi(x_i) | \psi(x_j) \rangle|^2
\end{equation}
where \( |\psi(x)\rangle = U_\phi(x)|0\rangle^{\otimes n} \) is the quantum state representing the classical data point \( x \), and \( U_\phi(x) \) is a data-encoding unitary circuit called the \textit{feature map}.

This fidelity-based kernel is then used to construct the \textit{quantum kernel matrix}, analogous to the classical kernel matrix. Once this matrix is computed—typically using the \textit{compute-uncompute} method on a quantum simulator or device—the remainder of the QSVM pipeline proceeds classically, using a conventional optimization solver based on the dual formulation of the SVM problem:
\begin{equation}
    \max_{\alpha} \sum_{i=1}^N \alpha_i - \frac{1}{2} \sum_{i,j=1}^N \alpha_i \alpha_j y_i y_j K(x_i, x_j)
\end{equation}
subject to
\begin{equation}
    \sum_{i=1}^N \alpha_i y_i = 0, \quad 0 \leq \alpha_i \leq C, \quad \forall i = 1, \ldots, N
\end{equation}
where \( \alpha_i \) are the Lagrange multipliers, \( y_i \) are the class labels, \( K(x_i, x_j) \) is the kernel function (quantum in the case of QSVM), and \( C \) is the regularization parameter.

QSVM thus blends quantum computation (for kernel evaluation) with classical optimization, aiming to exploit the expressive power of quantum states to better separate data in high-dimensional feature spaces.

Recent studies \cite{schuld2019quantum} have demonstrated that quantum kernels can outperform classical kernels on certain datasets by capturing richer structures through entanglement and interference. These theoretical advantages make QSVM a promising candidate for high-stakes classification tasks such as in medical diagnostics.

\subsubsection{Quantum Feature Maps}

\begin{figure}[H]
    \centering
    \includegraphics[width=0.5\textwidth]{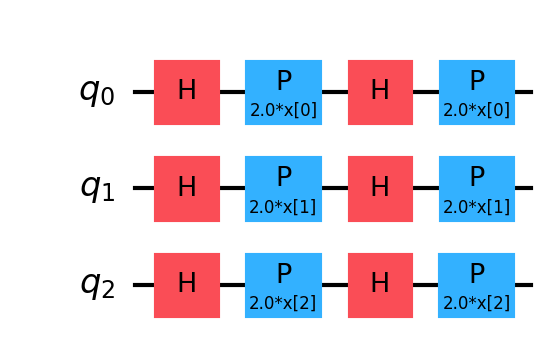}
    \caption{Quantum Circuit for ZFeatureMap}
    \label{fig:zfeaturemap}
\end{figure}

\begin{figure}[H]
    \centering
    \includegraphics[width=0.9\textwidth]{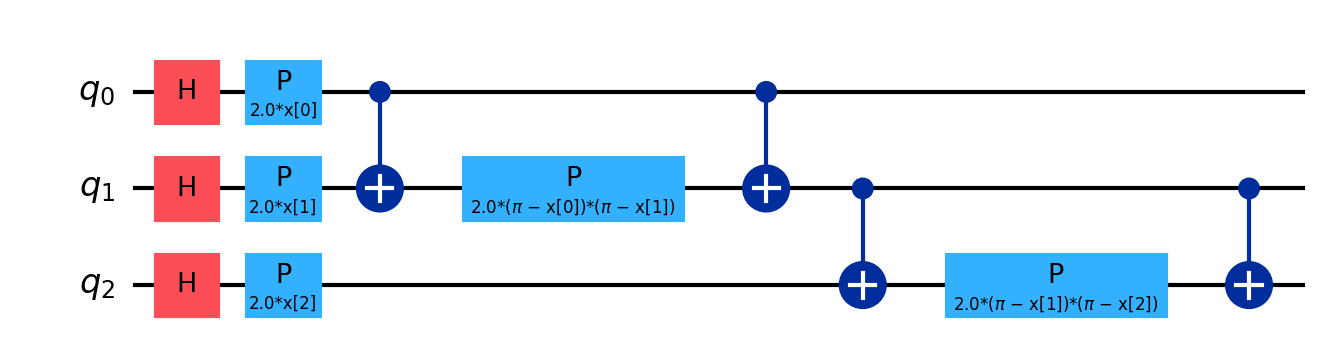}
    \caption{Quantum Circuit for ZZFeatureMap}
    \label{fig:zzfeaturemap}
\end{figure}

\begin{figure}[H]
    \centering
    \includegraphics[width=0.9\textwidth]{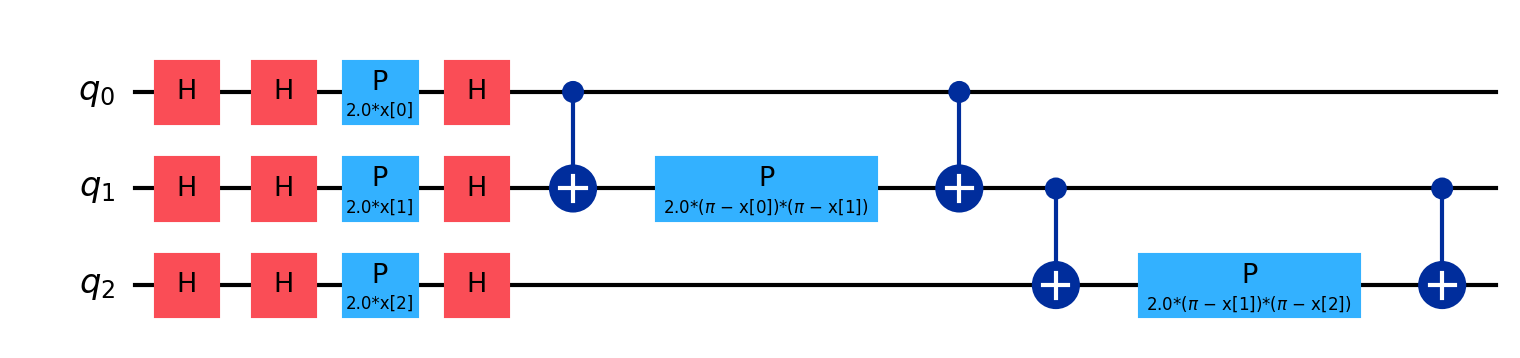}
    \caption{Quantum Circuit for PauliFeatureMap}
    \label{fig:paulifeaturemap}
\end{figure}

Quantum feature maps play a crucial role in Quantum Support Vector Machines (QSVM), as they define how classical data is embedded into quantum states. These feature maps influence the expressivity and noise sensitivity of the quantum model. In this subsection, we focus on three primary types of feature maps: ZFeatureMap, ZZFeatureMap, and PauliFeatureMap \cite{suzuki2024quantum, saad2022effect, bartkiewicz2019experimental}. The quantum circuits corresponding to these feature maps are shown in Figures~\ref{fig:zfeaturemap}, \ref{fig:zzfeaturemap}, and \ref{fig:paulifeaturemap}, respectively.\\

The \textbf{ZFeatureMap} is a Pauli-based feature map that applies single-qubit rotations around the Z-axis \cite{hafidi2023when}. While it does not involve entanglement, which limits its capacity to capture feature correlations, it is simple and efficient to implement, making it well-suited for datasets where features are assumed to be uncorrelated.

Though the theoretical formulation of the ZFeatureMap includes only \( R_z(2x_k) \) operations, in practice it is often preceded by Hadamard gates on each qubit to enhance the expressivity of the state. The unitary encoding is given by:

\begin{equation}
    U_{\phi}(\vec{x}) = \exp\left(i \sum_{k=1}^{n} 2x_k Z_k \right)
\end{equation}

where \( x_k \) is the \( k \)-th classical input and \( Z_k \) is the Pauli-Z operator acting on the \( k \)-th qubit.\\

The \textbf{ZZFeatureMap} introduces pairwise interactions between qubits using entanglement through the \( Z \otimes Z \) operation. It is better suited for correlated data and has higher expressive capacity than the ZFeatureMap, but is more sensitive to quantum noise. The theoretical formulation is:

\begin{equation}
    U_{\text{ZZ}}(\vec{x}) = \prod_{j=1}^{n} e^{ix_j Z_j} \cdot \prod_{j<k} e^{ix_j x_k Z_j \otimes Z_k}
\end{equation}

This entanglement allows the feature map to encode second-order feature interactions, which are crucial for tasks such as cancer diagnosis and fraud detection.\\

The \textbf{PauliFeatureMap} encodes classical data using rotations around all three Pauli axes (X, Y, Z) and includes entanglement terms. It is defined as:
\begin{equation}
    U_{\text{Pauli}}(\vec{x}) = \prod_{j=1}^{n} e^{-ix_j X_j} e^{-ix_j Y_j} e^{-ix_j Z_j} \cdot \prod_{j<k} e^{-ix_j x_k Z_j \otimes X_k}
\end{equation}
This map is the most expressive but also the most susceptible to noise due to the use of gates that flip the computational basis states, such as \( X \) and \( Y \).\\

The choice of an appropriate feature map depends on two major factors: the nature of the data and the noise resilience of the quantum device.

\begin{itemize}
    \item \textbf{Thermal Relaxation Noise:} ZFeatureMap is relatively resistant but can suffer from phase loss. ZZFeatureMap is more sensitive due to entanglement. PauliFeatureMap is highly sensitive to bit-flip noise introduced by \( X \) and \( Y \) gates.
    \item \textbf{Data Type:} ZFeatureMap is best for uncorrelated features and benefits from rapid execution. ZZFeatureMap is ideal for pairwise correlated features, albeit with higher computational cost. PauliFeatureMap, while powerful, requires careful noise mitigation strategies due to its complexity.
\end{itemize}
\subsection{Evaluation Metrics}

In this study, several key performance metrics were used to evaluate the effectiveness of the classification models, namely \textbf{accuracy}, \textbf{precision}, \textbf{recall}, \textbf{specificity}, and \textbf{F1-score}. These metrics provide a comprehensive understanding of how well the models perform in distinguishing between the two classes, which is especially important in healthcare applications where a correct classification is critical.

\textbf{Accuracy} is one of the most straightforward evaluation metrics, defined as the ratio of correctly classified instances to the total number of instances. It is given by:

\[
\text{Accuracy} = \frac{\text{True Positives} + \text{True Negatives}}{\text{True Positives} + \text{True Negatives} + \text{False Positives} + \text{False Negatives}}
\]

where:
\begin{itemize}
    \item \textit{True Positives} is the number of instances correctly classified as positive.
    \item \textit{True Negatives} is the number of instances correctly classified as negative.
    \item \textit{False Positives} is the number of instances incorrectly classified as positive.
    \item \textit{False Negatives} is the number of instances incorrectly classified as negative.
\end{itemize}

While accuracy is useful in many cases, it can be misleading when dealing with imbalanced datasets, which is why additional metrics are often considered.

\textbf{Precision} evaluates the number of positive predictions made by the model that are actually correct. It is defined as:

\[
\text{Precision} = \frac{\text{True Positives}}{\text{True Positives} + \text{False Positives}}
\]

Precision is particularly important when the cost of a false positive is high. In medical applications, a false positive might result in unnecessary treatment, so maximizing precision is crucial.

\textbf{Recall}, also known as sensitivity or the true positive rate, measures the ability of the model to identify positive instances correctly. It is given by:

\[
\text{Recall} = \frac{\text{True Positives}}{\text{True Positives} + \text{False Negatives}}
\]

In the context of healthcare, recall is especially important because it indicates how well the model can identify true positive cases (e.g., identifying patients who are at risk of developing a certain disease). A higher recall means fewer false negatives, which is critical when missing a positive case could have severe consequences.

\textbf{Specificity}, or the true negative rate, measures the proportion of negative instances that are correctly identified. It is defined as:

\[
\text{Specificity} = \frac{\text{True Negatives}}{\text{True Negatives} + \text{False Positives}}
\]

This metric is particularly useful when the cost of a false positive is high, as it measures how well the model avoids incorrectly labeling negative instances as positive.

\textbf{F1-score} is the harmonic mean of precision and recall, providing a balance between the two. It is particularly useful when both false positives and false negatives carry similar importance. It is defined as:

\[
F1\text{-score} = 2 \times \frac{\text{Precision} \times \text{Recall}}{\text{Precision} + \text{Recall}}
\]

The F1-score is a crucial metric when the class distribution is imbalanced, as it provides a better measure of the model's performance in both identifying positive and negative instances.

These evaluation metrics provide a comprehensive assessment of the model’s performance and are particularly important in imbalanced datasets where accuracy alone might not be sufficient. Each metric captures a different aspect of the classification process, helping to ensure that the models perform well in detecting both positive and negative instances in the data \cite{Sokolova2006, Powers2011}.
\section{Experimental Results}

The experimental setup for this study was based on a publicly available lung cancer dataset retrieved from Kaggle~\cite{sendil2025lung}. This dataset consists of 309 patient records, each annotated with a binary label indicating the presence (1) or absence (0) of lung cancer. Among the collected samples, 270 correspond to positive cases and 39 to negative cases, highlighting a significant class imbalance. The features are primarily binary (Yes/No responses), except for the gender attribute (Male/Female) and age, which is a continuous variable. All binary categorical features were numerically encoded by mapping "Yes" to 1 and "No" to 0, while gender was encoded by mapping "M" to 1 and "F" to 0. The continuous age feature was standardized using z-score normalization, ensuring zero mean and unit variance, a common practice to enhance numerical stability during model training~\cite{normalization}. No features were discarded from the original dataset in order to maintain a consistent feature space across classical and quantum models. Additionally, the dataset exhibited no missing values, eliminating the need for imputation or further cleaning procedures. To address the substantial class imbalance, six balanced subsets were constructed. Each subset contains all 39 negative samples and 39 unique positive samples selected randomly without replacement, as summarized in Table~\ref{tab:dataset_summary}. This strategy ensures that model evaluation is performed over multiple balanced configurations without overlap, preserving data integrity and minimizing bias—a critical consideration in healthcare-related machine learning tasks~\cite{imbalance_healthcare}.

\begin{table}[H]
\centering
\caption{Summary of the lung cancer dataset used in this study.}
\label{tab:dataset_summary}
\begin{tabular}{|c|c|}
\hline
\textbf{Characteristic} & \textbf{Details} \\ \hline
Total Samples & 309 \\ \hline
Positive Cases (y = 1) & 270 \\ \hline
Negative Cases (y = 0) & 39 \\ \hline
Number of Features & 15 (14 binary, 1 continuous) \\ \hline
Missing Values & None \\ \hline
Data Balancing Strategy & 6 balanced subsets (each 78 samples) \\ \hline
Feature Encoding & Binary mapping, Gender mapping, Age standardization \\ \hline
\end{tabular}
\end{table}

\begin{figure}[H]
\centering
\includegraphics[width=0.9\textwidth]{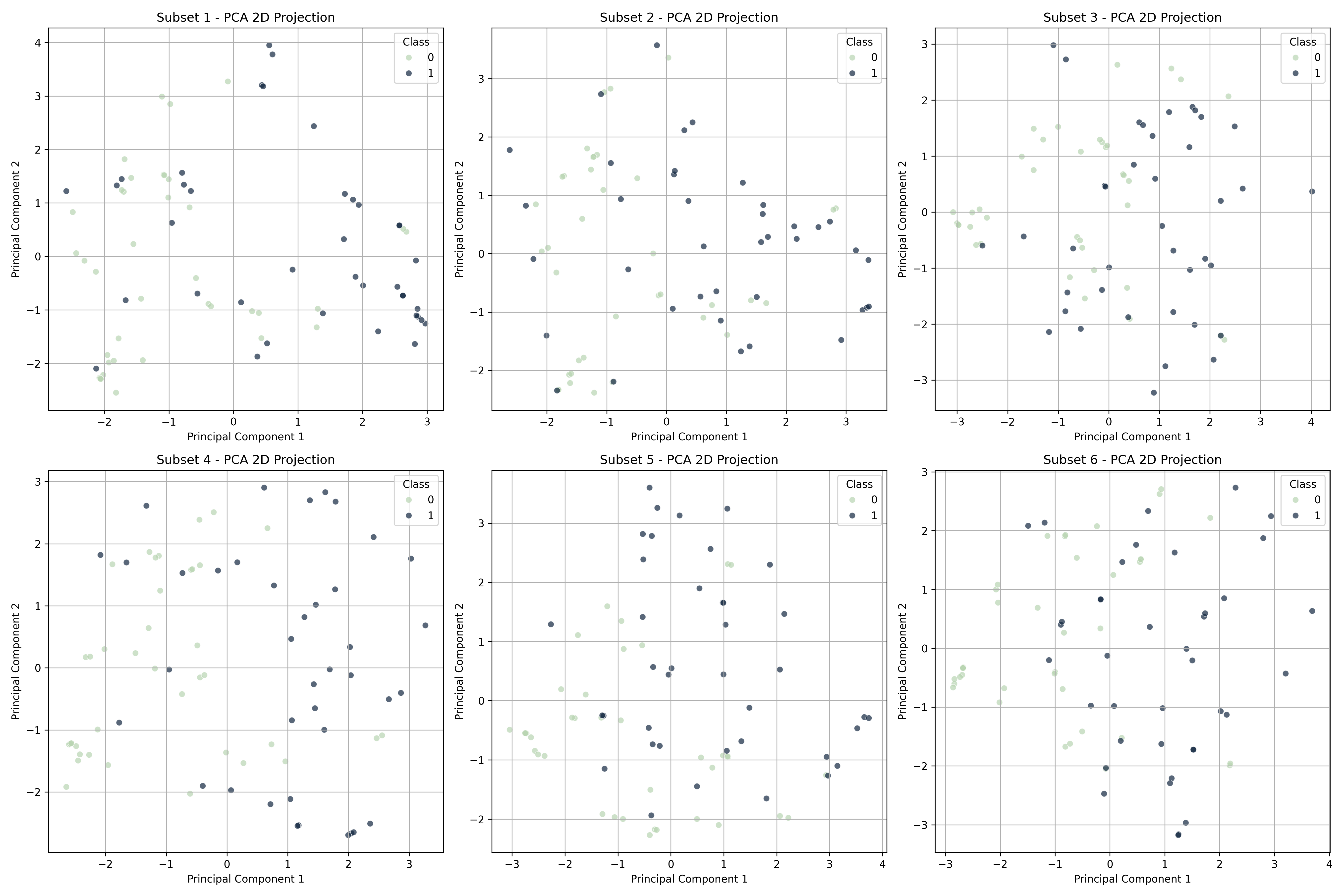}
\caption{PCA visualization of the six balanced subsets. Each plot represents a two-dimensional projection of the samples, colored by class label.}
\label{fig:pca_subsets}
\end{figure}

To better understand the structure and variability of the data within each balanced subset, a Principal Component Analysis (PCA)~\cite{jolliffe2002principal} was performed. PCA is widely used for dimensionality reduction and visualization in machine learning, enabling the projection of high-dimensional data into two principal components while preserving as much variance as possible~\cite{abdi2010principal}. Figure~\ref{fig:pca_subsets} illustrates the two-dimensional PCA projections for each of the six subsets. Each plot displays the separation between positive and negative classes, providing a visual confirmation that despite balancing, some overlap remains between classes. Such overlaps emphasize the classification challenge and justify the need for sophisticated modeling approaches like quantum-enhanced classifiers.

In order to establish a baseline for evaluating quantum models, we first conducted a comparative analysis between classical Support Vector Machines (SVM) and Quantum Support Vector Machines (QSVM). For the classical SVM models, three different kernel functions were tested: linear, polynomial, and radial basis function (RBF). The evaluation was performed across the six balanced subsets described previously, and the results were averaged to ensure robustness. The performance metrics considered include accuracy, precision, recall, specificity, and F1-score, which provide a comprehensive view of the classification effectiveness. For the linear kernel, SVM achieved an accuracy of 0.91, precision of 0.93, recall of 0.90, specificity of 0.92, and F1-score of 0.90. For the polynomial kernel, the results improved with an accuracy of 0.94, precision of 0.98, recall of 0.90, specificity of 0.98, and F1-score of 0.93. The radial basis function (RBF) kernel provided the best results, with an accuracy of 0.95, precision of 0.98, recall of 0.92, specificity of 0.98, and F1-score of 0.95.

For the quantum approach, QSVM models were trained using three different quantum feature maps: ZFeatureMap, ZZFeatureMap, and PauliFeatureMap. The PauliFeatureMap was configured with Pauli rotations on the $Z$, $X$ and $Y$ axes, enabling the encoding of complex feature correlations. Each quantum model was evaluated on the same six balanced subsets to ensure a fair comparison with its classical counterpart. The results for these feature maps are summarized as follows: for ZFeatureMap, the model achieved an accuracy of 0.88, precision of 0.95, recall of 0.81, specificity of 0.94, and F1-score of 0.86; for ZZFeatureMap, the results were an accuracy of 0.91, precision of 0.87, recall of 0.96, specificity of 0.85, and F1-score of 0.91; and for PauliFeatureMap, the model achieved a consistent performance of 0.96 across all metrics, including accuracy, precision, recall, specificity, and F1-score.

\begin{figure}[ht]
    \centering
    \includegraphics[width=0.8\textwidth]{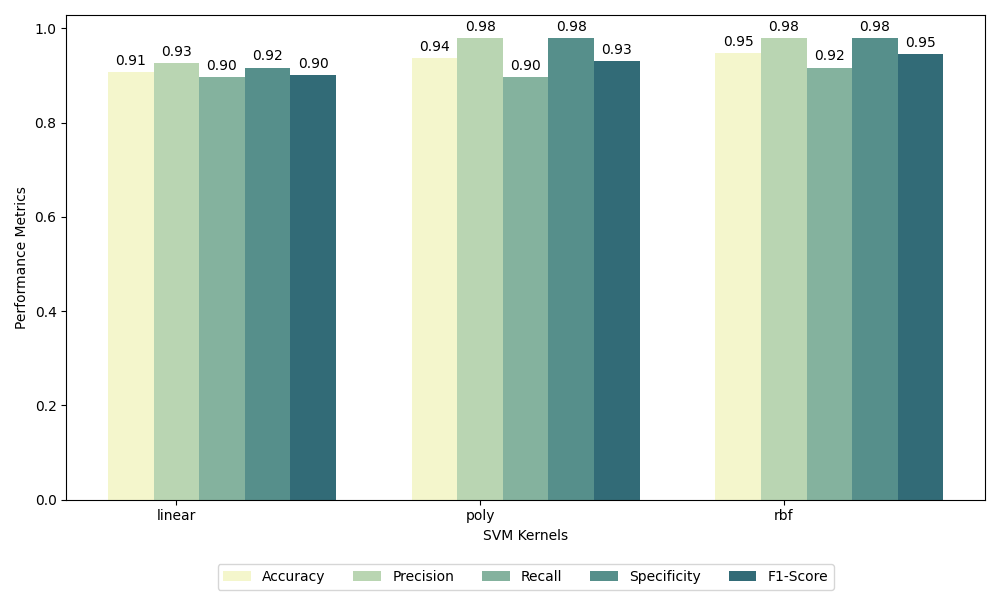}
    \caption{Grouped bar chart showing the average performance metrics of SVM models across different kernels (linear, polynomial, RBF) over six balanced subsets.}
    \label{fig:svm_grouped_bar}
\end{figure}

\begin{figure}[ht]
    \centering
    \includegraphics[width=0.8\textwidth]{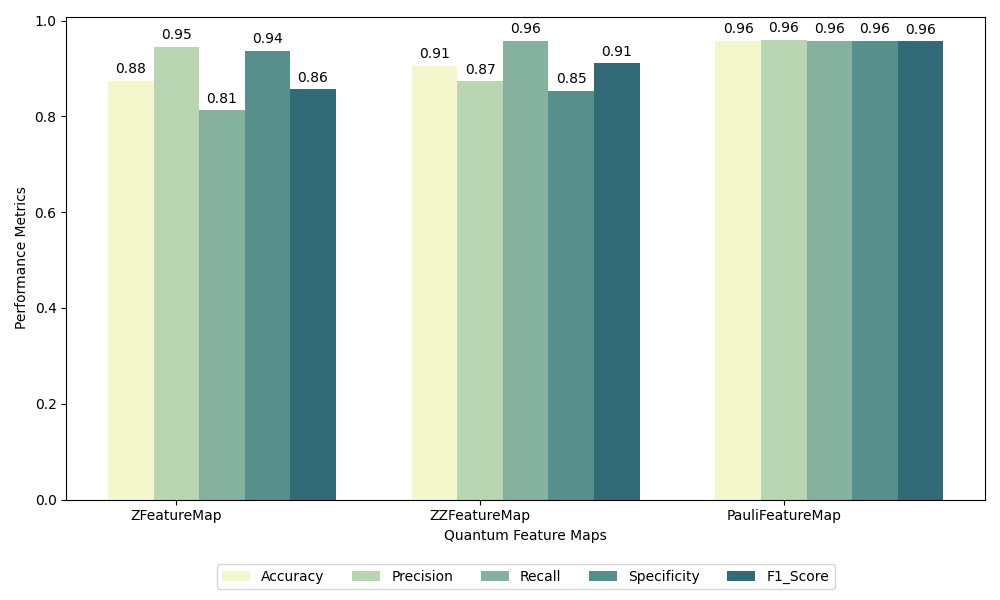}
    \caption{Grouped bar chart showing the average performance metrics of QSVM models across different quantum feature maps (ZFeatureMap, ZZFeatureMap, PauliFeatureMap with Z, X and Y rotations) over six balanced subsets.}
    \label{fig:qsvm_grouped_bar}
\end{figure}

The comparative results are summarized in two grouped bar charts. The first chart presents the average performance of classical SVM models across the different kernel types, while the second chart illustrates the average performance of QSVM models across the three feature maps. Figures \ref{fig:svm_grouped_bar} and \ref{fig:qsvm_grouped_bar} provide a visual summary of these findings, highlighting the variations in model behavior depending on the choice of kernel or quantum feature map.

Building upon the preliminary comparison between classical SVM and Quantum Support Vector Machine (QSVM), we now shift our attention to a more in-depth analysis of QSVM's performance across the six balanced subsets. In this section, we focus specifically on the impact of different quantum feature maps—namely, ZFeatureMap, ZZFeatureMap, and PauliFeatureMap—on QSVM's ability to classify the lung cancer dataset. The performance of QSVM is evaluated using five key metrics: accuracy, precision, recall, specificity, and F1-score, all of which are presented in Table \ref{tab:qsvm_results}.

The results provide insight into how each feature map influences the overall classification performance of QSVM. The three feature maps represent different quantum encoding schemes, each of which can capture and utilize the data's quantum characteristics in distinct ways. As detailed in the following table, we compare the performance of QSVM models across the six subsets, with each subset being evaluated using the three feature maps.

\begin{table}[h!]
    \centering
    \begin{tabular}{|c|c|c|c|c|c|c|}
    \hline
    \textbf{Feature Map} & \textbf{Subset} & \textbf{Accuracy} & \textbf{Precision} & \textbf{Recall} & \textbf{Specificity} & \textbf{F1-Score} \\
    \hline
    \multirow{6}{*}{\textbf{ZFeatureMap}} & 1 & 0.875 & 0.875 & 0.875 & 0.875 & 0.875 \\
    & 2 & 0.875 & 1.000 & 0.750 & 1.000 & 0.857 \\
    & 3 & 0.875 & 0.800 & 1.000 & 0.750 & 0.889 \\
    & 4 & 0.875 & 1.000 & 0.750 & 1.000 & 0.857 \\
    & 5 & 1.000 & 1.000 & 1.000 & 1.000 & 1.000 \\
    & 6 & 0.750 & 1.000 & 0.500 & 1.000 & 0.667 \\
    \hline
    \multirow{6}{*}{\textbf{ZZFeatureMap}} & 1 & 1.000 & 1.000 & 1.000 & 1.000 & 1.000 \\
    & 2 & 0.875 & 0.875 & 0.875 & 0.875 & 0.875 \\
    & 3 & 0.875 & 0.8000 & 1.000 & 0.750 & 0.889 \\
    & 4 & 0.937 & 0.889 & 1.000 & 0.875 & 0.941 \\
    & 5 & 0.875 & 0.800 & 1.000 & 0.750 & 0.889 \\
    & 6 & 0.875 & 0.875 & 0.875 & 0.875 & 0.875 \\
    \hline
    \multirow{6}{*}{\textbf{PauliFeatureMap}} & 1 & 0.875 & 0.875 & 0.875 & 0.875 & 0.875 \\
    & 2 & 0.937 & 1.000 & 0.875 & 1.000 & 0.933 \\
    & 3 & 0.937 & 0.889 & 1.000 & 0.875 & 0.941 \\
    & 4 & 1.000 & 1.000 & 1.000 & 1.000 & 1.000 \\
    & 5 & 1.000 & 1.000 & 1.000 & 1.000 & 1.000 \\
    & 6 & 1.000 & 1.000 & 1.000 & 1.000 & 1.000 \\
    \hline
    \end{tabular}
    \caption{Performance of QSVM across the 6 balanced subsets using three different quantum feature maps: ZFeatureMap, ZZFeatureMap, and PauliFeatureMap. The metrics shown include accuracy, precision, recall, specificity, and F1-score for each subset under the corresponding feature map.}
    \label{tab:qsvm_results}
\end{table}

From the results shown in Table \ref{tab:qsvm_results}, we observe that the PauliFeatureMap consistently yields the highest performance across all subsets, achieving perfect scores in Subsets 4, 5, and 6. This suggests that PauliFeatureMap is particularly effective at capturing the complex feature relationships required for accurate classification, leading to perfect recall, precision, and F1-score in these subsets.

The ZZFeatureMap, while demonstrating solid performance, shows some variability in its results. It achieves perfect scores for Subset 1, Subset 5, and Subset 6, but in other subsets, its performance drops slightly, particularly in recall. This variability indicates that while ZZFeatureMap is effective, its performance can be sensitive to the characteristics of the data in each subset.

The ZFeatureMap also performs well but generally yields lower results in comparison to the other two feature maps. It is most consistent in terms of performance, achieving stable accuracy across the subsets, but its recall and precision tend to be lower, especially in Subsets 2, 3, and 6. This suggests that ZFeatureMap may not be as well-suited to capturing complex patterns in the data compared to ZZFeatureMap and PauliFeatureMap.

Overall, these results demonstrate the importance of selecting the right feature map for quantum machine learning tasks. The PauliFeatureMap appears to be the most effective feature map for this task, particularly in subsets where the data may involve more complex relationships between the features. However, the ZZFeatureMap and ZFeatureMap also show promise, with ZZFeatureMap achieving strong results in several subsets. The choice of feature map plays a crucial role in the QSVM's classification performance, and further experiments could explore whether additional feature maps could lead to even better results.

\section{Discussion}

In this study, we examined the role of quantum feature maps—ZFeatureMap, ZZFeatureMap, and PauliFeatureMap—in Quantum Support Vector Machines (QSVM), particularly focusing on their structural differences, susceptibility to quantum noise, and data suitability.

Each feature map presents a trade-off between circuit complexity, expressivity, and robustness to quantum decoherence. The \textbf{ZFeatureMap} applies only single-qubit $Z$-rotations, resulting in shallow circuits with no entanglement. This simplicity makes it less sensitive to hardware noise and faster to execute, which is beneficial in near-term quantum devices. However, due to the absence of entanglement, it is best suited for datasets with uncorrelated features. Moreover, it begins to lose phase coherence under thermal relaxation, which causes high-probability states to degrade, thereby affecting classification performance in more complex tasks.

The \textbf{ZZFeatureMap} builds upon the ZFeatureMap by adding pairwise entanglement through controlled-ZZ interactions. This structure enables it to capture dependencies between features, making it more appropriate for datasets with pairwise correlations. As a result, it tends to achieve better classification accuracy than the ZFeatureMap when feature interactions are significant. Nevertheless, the entanglement increases the circuit depth and sensitivity to thermal noise. Thermal relaxation in ZZFeatureMap not only causes phase decay as seen in the ZFeatureMap but amplifies the effect due to entanglement, leading to higher degradation in quantum state fidelity.

The \textbf{PauliFeatureMap} generalizes feature encoding further by using a combination of Pauli gates ($X$, $Y$, $Z$), introducing both expressivity and complexity. Unlike the ZZFeatureMap, it often avoids entanglement but uses basis-changing operations that make it highly sensitive to bit-flip errors. This sensitivity arises because $X$ and $Y$ gates alter the computational basis, potentially leading to the emergence of new, unlikely states under thermal noise. While powerful in capturing intricate data patterns in simulations, the PauliFeatureMap shows higher variance in performance and may not generalize well on noisy intermediate-scale quantum (NISQ) hardware. Notably, it is still better suited for uncorrelated datasets and offers fast implementation due to limited entanglement, albeit with deeper single-qubit operations.

In summary, our results suggest that:
\begin{itemize}
    \item \textbf{ZFeatureMap} is most resilient to hardware noise and suitable for simpler, uncorrelated data, though limited in capturing complex relationships.
    \item \textbf{ZZFeatureMap} strikes a balance between expressivity and practicality, being ideal for correlated data but moderately affected by noise due to entanglement.
    \item \textbf{PauliFeatureMap} is the most expressive in theory but suffers from high noise sensitivity, particularly bit flips, due to basis-altering operations.
\end{itemize}

Choosing an appropriate feature map must account for both the nature of the dataset and the physical limitations of the quantum device. Future work may explore adaptive feature maps or hybrid classical-quantum encodings that better leverage the benefits of each approach while minimizing their drawbacks.

\section{Conclusion}

This research explored the impact of quantum feature map design on the performance of Quantum Support Vector Machines (QSVM) in binary classification tasks. By focusing on three commonly used feature maps—ZFeatureMap, ZZFeatureMap, and PauliFeatureMap—we demonstrated how each architecture uniquely influences model behavior in terms of expressivity, noise resilience, and data suitability. Our findings revealed that the ZFeatureMap is well-suited for uncorrelated datasets due to its simple structure and reduced sensitivity to noise, though it may suffer from phase relaxation. The ZZFeatureMap, incorporating quantum entanglement, showed improved performance on correlated data but exhibited increased vulnerability to thermal relaxation errors due to its deeper circuit. In contrast, the PauliFeatureMap, while theoretically more expressive, was most affected by noise—particularly bit-flip errors—because of its use of non-diagonal gates that alter the computational basis. These results highlight the critical role of feature map selection in optimizing QSVM performance and emphasize the importance of aligning quantum circuit design with both data structure and hardware limitations for practical quantum machine learning applications.

\bibliographystyle{unsrt}
\bibliography{references}

\end{document}